\documentclass[a4paper]{jpconf}
\usepackage{graphicx}
\begin{document}
\title{  
  The design of the time-of-flight system for MICE} 
\begin{center}
\vskip 0.5cm
{\large MICE Collaboration}
\end{center}
\author{M. Bonesini$^1$}

\address{$^1$ Sezione INFN Milano Bicocca, Dipartimento di Fisica 
G. Occhialini, \\
Piazza Scienza 3, Milano, Italy}

\ead{maurizio.bonesini@mib.infn.it}

\begin{abstract}
The international Muon Ionization Cooling Experiment (MICE) will carry
out a systematic investigation of ionization cooling of a muon beam.
As the emittance measurement will be done on a particle-by-particle
basis, a sophisticated beam instrumentation is needed to measure 
particle coordinates and timing vs RF. The 
MICE time-of-flight system will measure timings with a resolution better than 70 ps 
per plane, in a harsh environment due to high particle rates, fringe
magnetic fields and electron backgrounds from RF dark noise. 
\end{abstract}
\begin{figure}
\vskip -1cm
\begin{center}
\includegraphics[width=.70\linewidth]{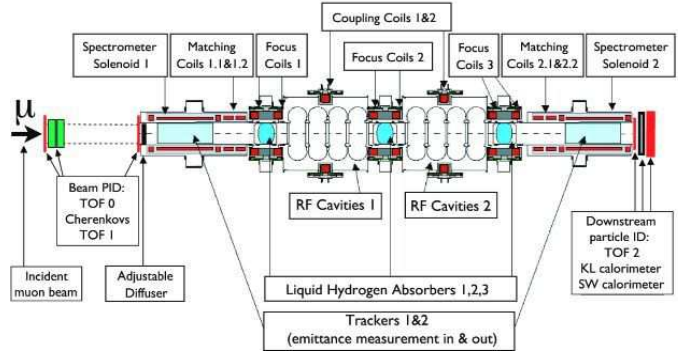}
\end{center}
\caption{2-D layout of the MICE experiment at RAL. 
The beam enters from the left. The cooling section
is put between two magnetic spectrometers and two TOF stations
(TOF1 and TOF2)
to measure particle parameters.}
\label{fig:mice}
\end{figure}

The neutrino factory ($\nu F$)~\cite{kosharev} is a muon storage ring with long straight 
sections, where decaying muons produce collimated neutrino beams of well defined
composition and high intensity. 
Several $\nu$F designs have been proposed, such as the ones of references
\cite{US2,cern}. 
The physics program at a neutrino factory is very rich and 
includes long-baseline $\nu$
oscillations, short-baseline $\nu$ physics and slow muon physics \cite{physrep}.
\noindent
The physics performances of a Neutrino Factory depend not only on its
clean beam composition ($50 \% \nu_{e}, 50 \% \overline{\nu}_{\mu}$ for
the $\mu^{+} \mapsto \overline{\nu}_{\mu} \nu_e e^{+}$ case), but
also on the available beam intensity. The cooling of muons 
(accounting for $ \sim 20 \%$ of the final costs of the factory)
is thus compulsory,
increasing the performances up to a factor 10. Due to the short 
muon lifetime ($2.2 \ \mu$s),
 novel methods such as the
ionization cooling, proposed more than 20 years ago by A.N. Skrinsky \cite{skrinsky}, must be used. 
Essentially the cooling of the transverse phase-space coordinates of a muon
beam can be accomplished by passing it through an energy-absorbing material
and an accelerating structure, both embedded within a focusing magnetic 
lattice. Both longitudinal and transverse momentum are lost in the absorber
while the RF-cavities restore only the longitudinal component. 
\noindent
The MICE experiment \cite{mice} at RAL aims at a
systematic study of a section of a cooling channel (see figure \ref{fig:mice} for a
layout).
A secondary muon beam from ISIS (140-240 MeV/c central momentum,
tunable between 1-12 $\pi \cdot $ mm rad input emittance) 
enters a cooling section after a diffuser. 
The 5.5 m long cooling section consists of three absorbers and eight RF
cavities encircled by lattice solenoids.
The cooling process will be studied by varying the relevant parameters, 
to allow the extrapolation to different cooling channel designs.
\section{The MICE TOF system}
\begin{figure}
\vskip -1.5cm
\begin{center}
\includegraphics[width=.35\linewidth]{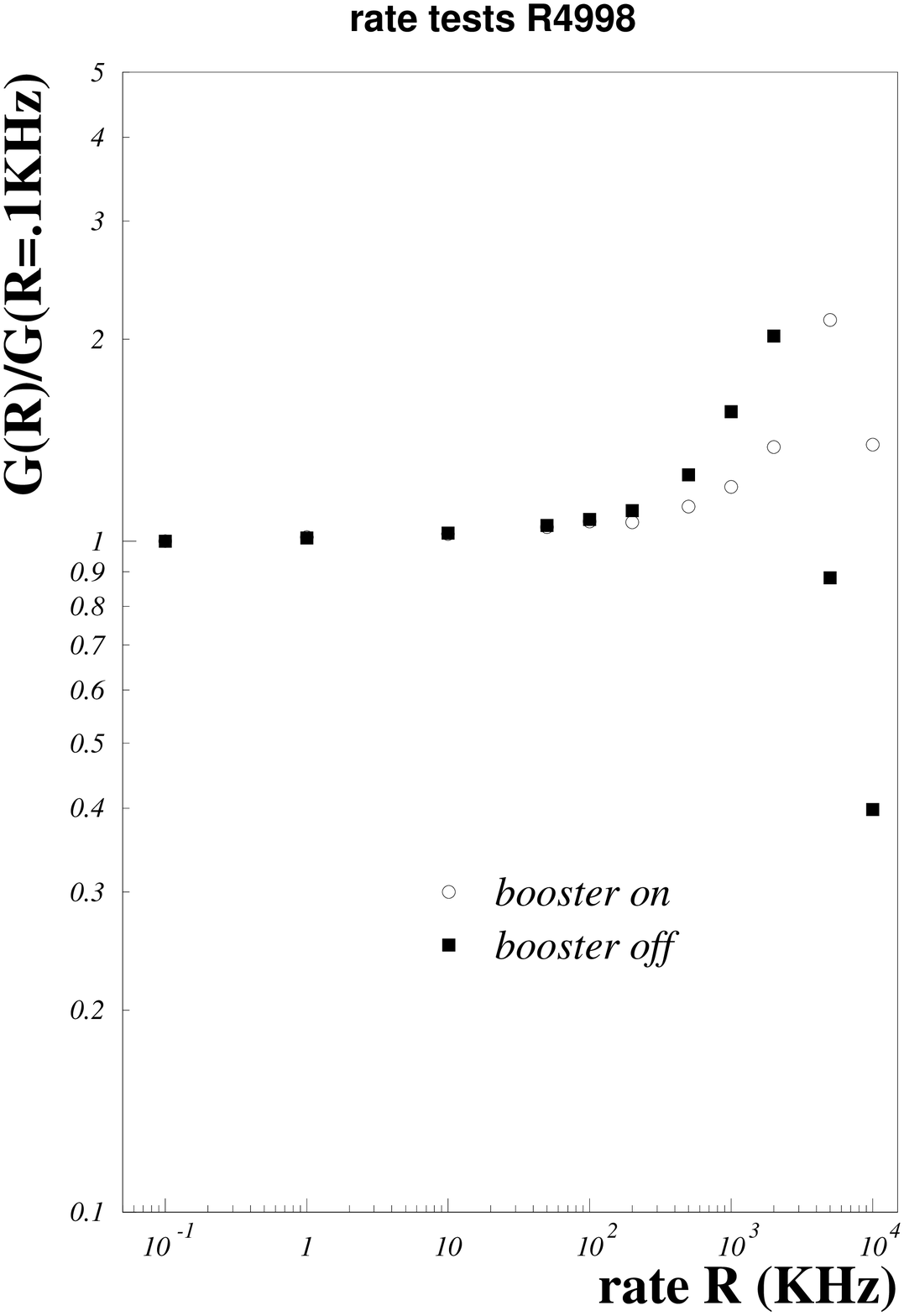}
\includegraphics[width=.35\linewidth]{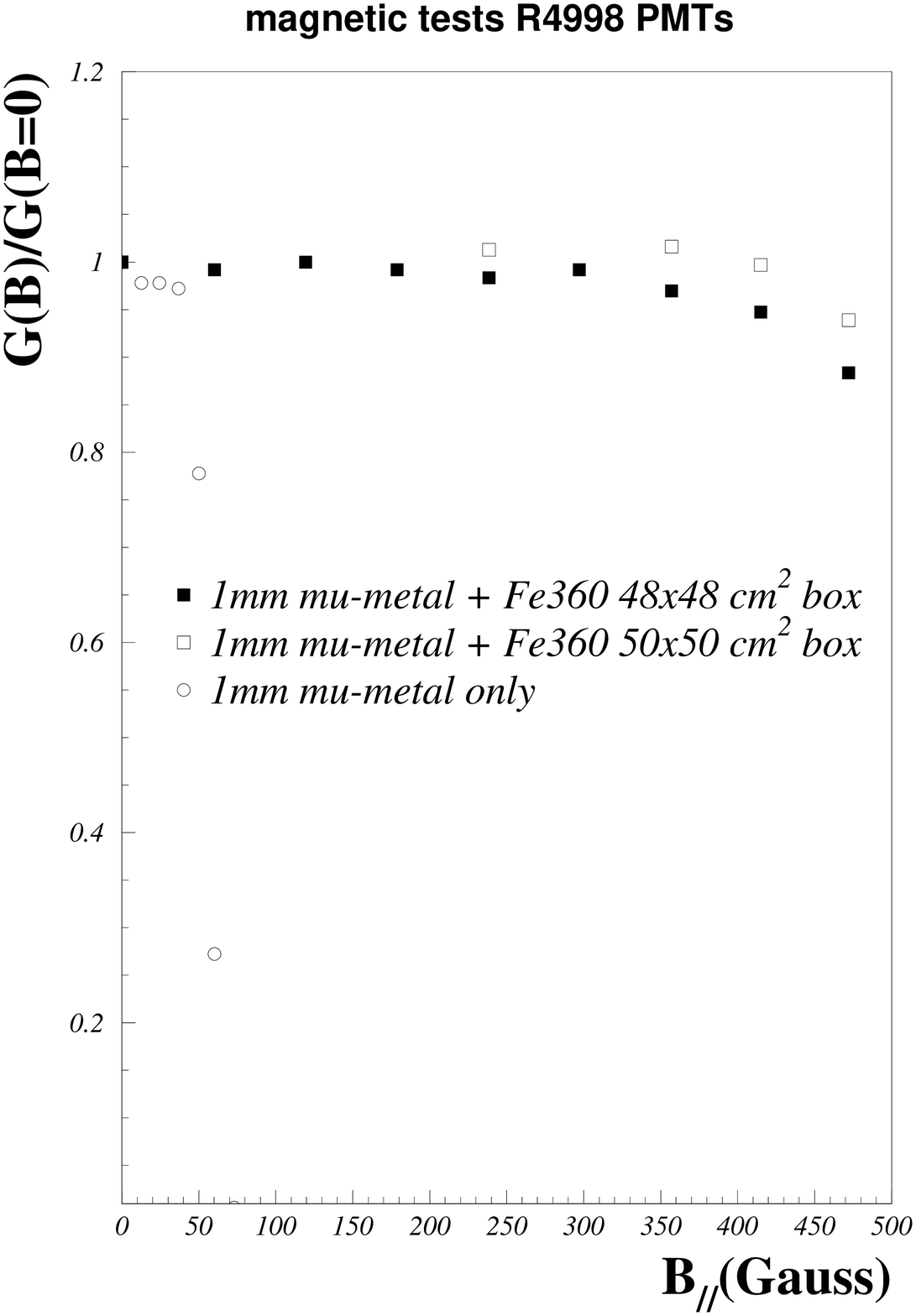}
\end{center}
\caption{Left panel: rate capability of a typical R4998 PMT, as a function of 
rate R at field B=0 G (measured P.H. in mV versus rate in KHz). 
Right panel: tests of shieldings for conventional R4998 Hamamatsu PMTs. 
The B field is along the PMT axis.}
\label{fig:pmt}
\end{figure}

Particles are measured before and after the cooling section
by two magnetic spectrometers complemented by TOF detectors. 
For each particle x,y,t, 
x'=dx/dz=$p_x/p_z$,y'=dy/dz=$p_y/p_z$, t'=dt/dz=$E/p_z$ coordinates 
are measured.
In this way, for an ensemble of N particles, the
input and output emittances are measured  with high precision ($0.1 \%$).  
Conventional multiparticle methods cannot be used, due to the correlation
between the six particle coordinates, induced by the presence of 
solenoidal magnetic fields.
\noindent
The driving design criteria of the MICE detector system~\cite{frascati}
 are robustness, 
in particular of the tracking detectors, to sustain the severe 
background conditions in the vicinity of RFs and redundancy in PID 
in order to keep contaminations ($e, \pi$) below $1 \%$.
Precision timing measurements are required
to relate the time of a muon to the phase of the RF and simultaneously
for particle identification by time-of-flight (TOF).
A time resolution around 70 ps ($\sigma_t$) provides both effective ($99 \%$) 
rejection of beam pions and adequate ($5^0$) precision of the RF phase.
Particle identification is obtained upstream the first solenoid by two 
TOF stations (TOF0/TOF1) and a Cerenkov counter (CKV1).
Downstream the PID is obtained via a further TOF station (TOF2)
and a calorimeter, to separate muons from decay electrons. 
The TOF stations share a common design based on fast 1" scintillator counters
along X/Y directions (to increase measurement redundancy) read at both
edges by R4998 Hamamatsu photomultipliers~\footnote{1" linear focussed PMTs,
typical gain $G \sim 5.7 \times 10^6$ at B=0 Gauss, risetime 0.7 ns, TTS 
$\sim 160$ps}. While TOF0 planes cover a $ 40 \times 40 \ cm^2$ active area,
TOF1 and TOF2 cover respectively a $ 42 \times 42 \ cm^2$ and 
$60 \times 60 \ cm^2$ active area. The counter width is 4 cm in TOF0 and
6 cm in the following ones.   
All downstream
detectors and the TOF1 station must be shielded against stray magnetic
fields (up to 1000-1500 Gauss with a $\leq 400$ Gauss longitudinal component, 
depending on the design of the shielding plates after the spectrometer 
solenoids). Two options for the local TOF 1/2 shielding are under study: 
in one a double-sided shielding cage will contain fully the detector, 
aside an hole for beam,   while in the other individual
massive soft iron box PMTs shielding are under study \cite{ref_D0}.
While the first solution is more elegant and reduce the detector weight,
it gives complications for detector access and maintenance. 
Figure \ref{fig:pmt} shows some preliminary results for the shielding of
the most dangerous component of the B field, along the PMT axis, obtained
with simple mu-metal and mu-metal+a massive iron box shielding. 
   
Counter prototypes have been tested at the LNF $DA \Phi NE$ Beam Test Facility
(BTF) with incident electrons of $E_{beam}=200-350$ MeV to study the 
intrinsic counter time resolution. The frontend readout
used the baseline MICE choice for the TDC: a multihit/multievent CAEN 
V1290 TDC,  
in addition to a  CAEN V792 QADC (to be replaced in the experiment by 
CAEN V1724 FADC) for time-walk corrections. The PMT signal was splitted
by a passive splitter followed by a leading-edge discriminator before 
the TDC line. 
An intrinsic single counter resolution $\sim 45-60$ ps was obtained 
depending from beam conditions and the design of lightguides or the 
used scintillator
(Bicron BC404 or BC420~\footnote{ risetime 0.7 (0.5) ns, 
$\lambda^{max}_{emission}=408 (391)$ nm, $\lambda^{bulk}_{att}=160 (110)$ cm
for BC404 (BC420)}, Amcrys-H UPS95F).
In the same runs, assuming a gaussian fit for the pulse-height distribution
it was possible to estimate the number of photoelectrons per single impinging
electron ($N_{pe}$). From $(<R>/\sigma_R)^2$, where $<R>$ is the peak of the
gaussian and $\sigma_R$ its width, an estimate in the range 
200-300 p.e.  for BC420 was obtained, depending on the impact point.
Clearly, this estimation neglegts electronic noise and is affected
by the bad (good) scintillator-PMT coupling.

\noindent
The TOF stations must sustain a high incoming particle rate (up to 1.5 MHz
for TOF0).
PMTs rate capabilities were tested in laboratory with a dedicated
setup \cite{ref_laser} based on a fast laser. 
An home-made system based on a Nichia NDHV310APC violet laser diode and 
an AvetchPulse fast pulser (model AVO-9A-C laser diode driver, with 
$\sim 200$ ps risetime and a AVX-S1 output module) was used. This system
gave laser pulses at $\sim 409 $ nm, with a FWHM between $\sim 120$ ps and
$\sim 3$ ns (as measured with a 6GHZ 6604B Tek scope) and a max
repetition rate of 1 MHz. A typical R4998 PMT had a good
rate capability for signals comparable to an incident $\mu$ ($\sim \ 300 p.e.$)
up to $\sim 1$ MHz. The rate capability was increased by the use of active
bases or a booster on the last dynodes for the R4998 PMTs, as shown 
in figure \ref{fig:pmt}.    

\end{document}